\documentclass[%
 reprint,
 amsmath,amsfonts, amssymb
 aps,prl
]{revtex4-2}

\usepackage{graphicx,xcolor,subfig}
\usepackage{dcolumn}
\usepackage{bm}
\usepackage{hyperref}

\newcommand{\taurec}{\tau_{\rm rec}}
\newcommand{\tauturb}{\tau_{\rm turb}}
\newcommand{\Uin}{U_{\rm in}}
\newcommand{\Bin}{B_{\rm in}}
\newcommand{\lcs}{L_{\rm CS}}
\newcommand{\recrate}{\mathcal{R}}
\newcommand{\Erec}{E_{\rm rec}}
\newcommand{\Scrit}{S_{\rm crit}}

\def\apj{Astrophys.~J.}
\def\apjl{The Astrophysical Journal Letters}

\def\grl{Geophys.~Res.~Lett.}

\def\nat{Nature}

\def\pre{Phys.~Rev.~E}
\def\prl{Phys.~Rev.~Lett.}

\def\jgr{J.~Geophys.~Res.}

\begin{document}

\title{A transport-like approach to turbulence-mediated magnetic reconnection}

\author{Nuno F. Loureiro}
 \email{nflour@mit.edu}
\affiliation{Plasma Science and Fusion Center, Massachusetts Institute of Technology, Cambridge, MA 02139, U.S.A.}

\date{\today}

\begin{abstract}
A transport-like framework for the study of magnetic reconnection mediated by self-driven turbulence is proposed, based on timescale separation between the reconnection time and the characteristic timescale of the turbulent fluctuations which arise in the reconnection layer. 
We argue that the mean fields remain on MHD scales even in collisionless cases. 
These observations provide theoretical justification for an efficient computational approach to this problem, which we discuss.
\end{abstract}

\maketitle

\paragraph{Introduction.} 
Plasmas have the ability to irreversibly and non-trivially alter the topology of the magnetic fields that thread them. This nonlinear process --- known as magnetic reconnection~\cite{biskamp_2000,yamada_2010} --- underlies explosive events such as flares in astrophysical bodies~\cite{pettersen_1989,lyubarsky_2021}, as well as a wide variety of other phenomena, including the interaction between the magnetic fields of the Sun and the Earth~\cite{sonnerup_1981,burch_2016}. 
Reconnection is also thought to play an important role in plasma turbulence~\cite[e.g.][]{matthaeus_1986,Franci_2017,loureiro_2020,zhou_2020,schekochihin_MHD_turbulence_2022} and in the nonlinear development of many plasma instabilities~\cite[e.g.][]{Wesson_1990, nykyri_2001,leonard_2014,inchingolo_2018}.

As the understanding of reconnection matures, it is becoming increasingly clear that laminar reconnection sites (current sheets) are an exception, realizable only in systems with relatively small scale separation (low Lundquist number, or small ratio of system size to kinetic scales). 
More generally, reconnecting systems appear capable of driving their own turbulence; and, in turn, such turbulence becomes critical in defining the properties of the reconnecting system~\cite{daughton_2011,lapenta_2011,oishi_2015,huang_2016,beresnyak_2017,kowal_2017,yang_2020,kowal_2020,guo_magnetic_2021,zhang_efficient_2021,beg_2022}.

Understanding the interaction between the reconnecting field and the turbulence that it drives  --- particularly in realistic three-dimensional geometries where reconnection-driven turbulence can live up to its full potential --- is only just beginning, with progress severely limited by the extreme computational resources required.
In this Letter, we argue that a transport-like approach --- conceptually akin to how the problem of turbulent transport in fusion devices is usually approached~\cite[e.g.,][]{barnes_2010,Abel_2013} --- is justified and both enables the derivation of interesting results as well as yields a feasible and computationally efficient way forward in this problem. 

\paragraph{Mean-field decomposition.} 
As is standard practice in mean-field electrodynamics, we decompose fields (such as magnetic and velocity fields) into a mean and a fluctuating part: $\bm{Q}(\bm{r},t)=\bm{Q}_{0}(\bm{r},t)+\bm {q}(\bm{r},t)$, where $\bm{Q}_{0}=\overline{\bm{Q}}$ denotes the background, mean value of any field $\bm{Q}$ (with overbar representing an adequately defined spacial or temporal average, as appropriate),  and $\bm{q}$ is the fluctuating component (absent in the laminar limit).
The latter is assumed here to be driven by magnetic reconnection; i.e., as reconnection proceeds, it triggers instabilities (of the macroscopic sort, such as plasmoid or kink, and/or microscopic, such as mirror~\cite{alt_2019,winarto_2022}, Buneman~\cite{che_2010}, or others~\cite[e.g.][]{carter_2002,che_2011,li_2022}) which drive the fluctuations in the current sheet. 
Away from the current sheet, such fluctuations are assumed to be absent: there, the fields reduce to their mean values. 
Thus, the reconnection geometry we consider is one where the length $\lcs$ (in the outflow direction) is determined by the large scale variation of the mean (reconnecting) magnetic field $\bm{B}_0$. 
The scale of variation of the mean field across the current sheet, $a$, on the other hand, is dynamically set, and is to be determined; it effectively corresponds to a critical gradient that drives the fluctuation power required to reach a statistical steady state. 
We assume that $\lcs/a\gg 1$.

Essential to a mean-field decomposition is the (time- and/or length-) scale separation between mean and fluctuating quantities. 
We will argue below that spacial separation is not, in general, guaranteed. 
Timescale separation, however, is expected once a statistical steady-state is reached.
Indeed, the characteristic timescale of the fluctuations is the Alfv\'en time, $\tau_A\equiv\lcs/V_A$ (the advection time out of the sheet), or faster; whereas the mean fields evolve on the reconnection time, $\taurec=\recrate^{-1}\tau_A$, with $\recrate$ the normalized reconnection rate (a finite but small fraction of unity both in MHD and in collisionless cases). 
We shall denote the typical turbulence timescale as $\tauturb$ (which may be equal to $\tau_A$ in some cases, but not necessarily).
Therefore, mean fields, $\bm{Q}_{0}(\bm{r},t)$, are defined as a time average of the full fields over a time interval $\tau$ such that $\taurec\gg\tau\gg \tauturb$. 

Assuming that time-scale separation holds, what equations to solve for the mean fields depends on whether  $a$ is larger or smaller than the kinetic scales (the ion skin depth or the ion (sound) Larmor radius, as appropriate).
If larger, then the relevant equations are simply the well-known mean-field equations of MHD (see, e.g., Eqs. (4.2-4.3) of Ref.~\cite{biskamp_turbulence_2003} in the incompressible case). 
If, instead, $a$ is smaller than the ion kinetic scale, a richer description is needed.
We will argue, however, that this does not, in general, happen; i.e., that the scale of variation of the mean-fields, once a steady-state is reached, is larger than the kinetic scales  --- see discussion below. 
As for the fluctuating fields, one can use descriptions of varying degree of sophistication, from MHD to first-principles (Vlasov or PIC), as required by the specific problem at hand.

Finally, it is worth noting at the outset some key differences between the application of the mean-field approach to magnetic reconnection versus the kinematic dynamo, where it has been so successful~\cite{hughes_2018,rincon_2019,moffatt_2019}. 
First and foremost, the background magnetic field is obviously not small compared to the background flows; i.e., the kinematic approximation does not hold. 
Second, as we will discuss later, the background mean fields are inhomogeneous on scales that can be commensurate with those of the turbulent fluctuations; they are also anisotropic. 
And, third, the fluctuations cannot, in general, be assumed small --- on the contrary, they must be assumed to be Alfv\'enic with respect to the reconnecting (upstream) magnetic field.
These constraints strongly limit the ability to make progress on this problem (nonetheless, see Refs.~\cite{higashimori_2013,widmer_2019}).
However, it is both possible to derive a number of useful and interesting results using this approach, as well as devise an efficient computational framework that casts this as a self-consistent transport problem. 
We first analyze separately the MHD and the collisionless cases.

\paragraph{Strong guide-field reconnection in the MHD limit.}
We specialize for simplicity to guide-field reconnection in an incompressible plasma. There is a constant, homogeneous magnetic field in the direction perpendicular to the reconnection plane, $B_z \bm{\hat z}$. 
We assume $B_z\gg B_{\perp,0}$ and,
for concreteness, consider a reconnection geometry where $x$ and $y$ are, respectively, the inflow and outflow directions.
The global Lundquist number (defined with the reconnecting field) is $S=\lcs V_{A,y}/\chi_m\gg \Scrit\gg 1$, where $\chi_m$ is the magnetic diffusivity, and $\Scrit$ is a reference critical value required to trigger the plasmoid instability~\cite{loureiro_2007,bhattacharjee_2009,huang_2010,loureiro_2013, loureiro_2016} in the layer. 
Once the system settles down to a statistical steady-state, we expect a spectrum of highly anisotropic turbulent fluctuations whose outer scale in the $x$-direction is $a$, the width of the mean current sheet; i.e., the fluctuations define how broad the mean current sheet is. 

The equation for the mean electric field is
\begin{align}
\label{eq:mean_e_field}
    c\bm{E}_0 + \bm{U}_{\perp,0}\times \bm{B}_{\perp,0} +\overline{\bm{u}_\perp\times\bm{b}_\perp}=\eta c \bm{J}_0.
\end{align}
Away from the current sheet there are no fluctuations (and the resistive term is negligible) so Ohm's law yields simply $cE_0=\Uin\Bin$, with $\Uin$ and $\Bin$ the magnitudes of the upstream inflow velocity and reconnecting field (note that this would be true even if there were no timescale separation between mean fields and fluctuations; it requires only that the fluctuations be confined to the layer).

Seeking an approximate steady-state solution for the mean-fields implies that the mean electric field should be spatially constant (this is nothing but the standard Sweet-Parker argument~\cite{sweet_1958,parker_1957}).
Close to the center of the layer, where the (perpendicular) mean fields vanish (by symmetry), the electric field must then be set by the turbulent fluctuations (assuming that those are much larger than the resistive term, which must be the case for the problem to differ significantly from the standard (laminar) SP configuration).
Therefore, imposing (approximate) steady-state yields an expression for the reconnection electric field:
\begin{align}
    \label{eq:Erec}
    c \Erec\approx \text{const.}\approx \Uin\Bin\approx \left[ -\bm{\hat{z}}\cdot(\overline{\bm{u}_\perp\times\bm{b}_\perp})\right]_X,
\end{align}
where $[\cdots]_X$ means the expression is to be evaluated at the $X$ point of the mean fields. 
This relation between upstream quantities and the fluctuations in the layer can be directly tested in simulations, since all quantities are directly measurable. 
Its validity requires only that two hypotheses hold true: timescale separation between the evolution of the background fields and the fluctuations, and that a (quasi) steady-state is reached --- both of which seem reasonable to assume.
Although we have in mind 3D reconnecting systems whose layers host fully developed turbulence, these two hypotheses are also valid for the better understood (and simpler) case of 2D MHD stochastic plasmoid chain dynamics; so, it is worthwhile checking that Eq.~(\ref{eq:Erec}) is consistent with that understanding. The answer is that it is.
Indeed, in a plasmoid chain, the quantity $u_y b_x$ (evaluated along $x=0$) always has the same sign. 
Following Ref.~\cite{uzdensky_2010},
the expected value for $u_y$ is $V_{A,y}$, and $b_x\sim b_y w_x/w_y$, where $w_x,\, w_y$ are the typical plasmoid widths in the $x$ and $y$ directions.
The typical field in the plasmoids is the upstream field, $b_y\sim \Bin$, and $w_x/w_y\sim \recrate$.
Therefore, $u_y b_x\sim  V_{A,y} \recrate\Bin\sim \Uin \Bin$.

More generally, since $\Bin$ is assumed to be known, (\ref{eq:Erec}) is an equation for the inflow velocity $\Uin$ in terms of the fluctuations in the layer. 
If one assumes that 2D plasmoid chain physics extends to 3D, then $b_{y,\rm rms}\sim \Bin$, implying that
\begin{align}
    \label{eq:uin_mhd}
    \Uin\approx u_{x,\rm rms}. 
\end{align}
While perhaps somewhat obvious, this expression is a testable prediction. 
The fact that it matches the analytical result found in Appendix D.7.1 of Ref.~\cite{schekochihin_MHD_turbulence_2022} (Eq.~(D108)), arrived at via a different line of reasoning, and is in agreement with recent numerical results~\cite{yang_2020}, suggests that the mean field framework proposed here rests on solid grounds.

\begin{figure*}
    \centering
    \includegraphics[width=0.49\textwidth]{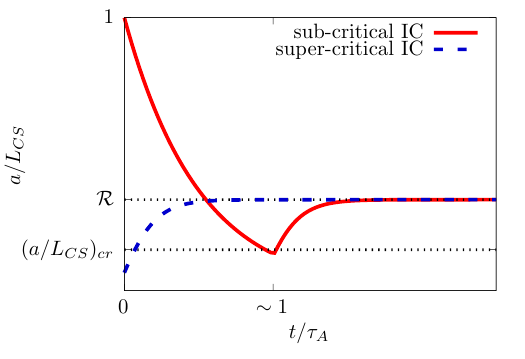}
    \includegraphics[width=0.49\textwidth]{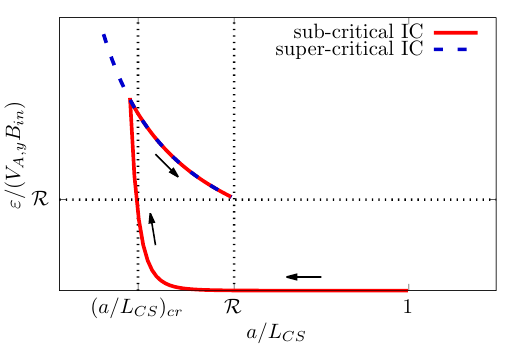}
    \caption{Conceptual evolution of a turbulent reconnecting system. Left: Aspect ratio of the mean current sheet as a function of time. Right: Normalized turbulent e.m.f. as a function of mean current sheet aspect ratio. The arrows indicate the time evolution of the system. In both plots, the bold (red) line indicates a case where the initial condition is sub-critical; the system evolves until it reaches a critical threshold for triggering an instability (e.g., plasmoid formation), and subsequently relaxes until a statistical steady-state is reached. The dashed (blue) line is instead the case where the initial aspect ratio of the background is super-critical and, therefore, strongly unstable.}
    \label{fig:plots}
\end{figure*}

In itself, Eq.~(\ref{eq:uin_mhd}) is not a solution to the problem: neither the upstream inflow velocity nor the outer-scale velocity fluctuations are \textit{a priori} known (indeed, determining $\Uin$, i.e., the reconnection rate, is the main goal). 
To address this, imagine that one initializes a SP-type reconnection layer (at $S\gg\Scrit$). 
Plasmoid turbulence will immediately ensue, broadening the mean layer~\cite{oishi_2015,huang_2016,beresnyak_2017,kowal_2017,yang_2020}; that is, the mean current sheet thickness and outflow velocity profile will be determined by the mean width of the fluctuations. 
If one now thinks of the stability of the mean profiles, one sees that if the layer were to broaden too much, the fluctuations would die out --- such a sheet would be stable to plasmoid formation --- and a steady-state would be impossible to maintain. 
The conclusion, then, is that the mean profiles have to relax and stay close to marginal stability: that is, an effective SP sheet with $a/\lcs\approx \Scrit^{-1/2}$; see Fig.~\ref{fig:plots}. In an approximately incompressible system, this is also the reconnection rate.
This is indeed the case in 2D MHD~\cite{uzdensky_2010,huang_2010,loureiro_2012}.
In 3D MHD there is less certainty on what $\Scrit$ is, though simulations suggest that it may not be significantly different than the 2D value~\cite{oishi_2015,loureiro_2016,beresnyak_2017}.
If that is the case, the mean reconnection rate would be $\approx 0.01$ --- in very reasonable agreement with existing numerical simulations~\cite{beresnyak_2017, kowal_2017, yang_2020}. 
Lastly, note that while this reconnection rate is ``fast'' in the usual sense --- it does not depend on resistivity --- it is small enough that time averaging of the fluctuations, as described above, is legitimate: $\tauturb=a/u_{x,a}\sim a/\Uin\sim \lcs/V_{A,y}\ll \taurec$.

\paragraph{Collisionless reconnection.}
Let us now consider
reconnection in weakly collisional plasmas.
While the range of possible instabilities driving turbulence in the current sheet is much greater in this case than in MHD, one might still expect tearing-driven (and potentially kink-unstable) flux ropes (plasmoids) to be among those, as suggested by available numerical studies~\cite[e.g.,][]{daughton_2011,guo_magnetic_2021,zhang_efficient_2021}.
Unlike other possible (intrinsically kinetic) instabilities, the structures that arise via these instabilities are larger scale --- and, as we will now conjecture, will be on MHD scales. 

Consider first the case of low plasma beta and strong guide field; and assume that a current sheet is forming with a rate $U_{\rm dr}/\lcs$ (which we take to be of the same order as the inverse Alfv\'en time, $V_{A,y}/\lcs$, such that $M_{\rm dr}\equiv U_{\rm dr}/V_{A,y}\sim 1$)~\cite{uzdensky_2016,tolman_2018}. 
One can then show that the growth rate of the most unstable tearing wavelength overcomes the current sheet formation rate when $a/\lcs\sim M_{\rm dr}^{-1/3} (d_e \rho_s/\lcs^2)^{1/3}$~\cite{sarto_2016,mallet_2020}\footnote{Assuming a Harris sheet profile for concreteness; other choices have no qualitative impact on the conclusion.}. 
At this moment of time, the current sheet width is still at MHD scales, $a>\rho_s$, if $\lcs/\rho_s > (\beta_e\, m_i/m_e)^{1/2} M_{\rm dr}$; this condition is very easy to satisfy.
One arrives at a similar condition for plasma beta of order unity (using two fluid tearing mode scalings~\cite{fitzpatrick_2004,fitzpatrick_2007}); in this case, $a>d_i$ requires $\lcs/d_i > (m_i/m_e)^{1/2}M_{\rm dr}$. 
Yet another example is that of (low beta, non-relativistic) pair plasmas~\cite{loureiro_2018}; there, the critical aspect ratio is $a/\lcs\sim M_{\rm dr}^{-1/3} (d_e\lcs)^{-2/3}\gg (d_e/\lcs)$.
Although we cannot prove this, we conjecture that this conclusion is more general than the specific examples just listed; i.e., that a current sheet forming in a large-scale, weakly collisional plasma will disrupt due to the tearing instability before its thickness reaches kinetic scales.

Once the plasmoid-driven turbulence sets in the layer, one expects that the reconnection rate will converge to a (time-averaged) constant $\recrate$: a small but finite fraction of the Alfv\'en rate (perhaps the canonical $0.1$ value~\cite{birn_2001,cassak_2017}).
Assuming compressibility effects are negligible, this ought to correspond to a mean current sheet whose aspect ratio is $a/\lcs\approx\recrate$. 
Now, recall that $\lcs$ is a macroscopic, system-size length and, thus, $\rho_s/\lcs\ll \recrate\ll 1$. 
So, regardless of whether the layer widens or contracts once the turbulence develops, the thickness of the mean layer, $a$, will have to be on MHD scales --- a conclusion which, from the point of view of the mean fields, is the only way to ensure that the reconnection rate $\recrate\approx a/L$ is independent of kinetic scales, as is suggested by multiple numerical studies~\footnote{In agreement with these ideas, we note that Beresnyak's simulations of 3D Hall MHD reconnection~\cite{Beresnyak_2018} develop a turbulent current layer which broadens well beyond the ion skin depth.}.

These considerations suggest that the mean field equations can still be those of MHD in this case, 
and that the first-principles Vlasov (or PIC) description is only required for computing the fluctuations in the layer.
Specifically, this implies that Eq.~(\ref{eq:mean_e_field}) applies both in the MHD and in the collisionless case; and so should Eq.~(\ref{eq:uin_mhd}).

\paragraph{Simplest possible subgrid model.}
While the framework that we have discussed so far requires one to rigorously compute the fluctuations in the layer, it may desirable, from a computational modelling perspective, to have a subgrid model for the fluctuations, such that only the (MHD) mean-field equations would have to be solved. 
The simplest such model is one where the turbulent e.m.f. in Eq.~(\ref{eq:mean_e_field}) is replaced with a turbulent resistivity. 
Dimensionally, this can be estimated as
\begin{equation}
    \label{eq:eta_turb_0}
    \eta_{\rm turb}\approx u_{x,a} a \approx \Uin a \approx V_{A,y} a^2/\lcs,
\end{equation}
where we have used Eq.~(\ref{eq:uin_mhd}).
One can simplify this further with the extra assumption that there is a critical aspect ratio below which the current sheet is unstable (e.g., around 0.01 in MHD):
\begin{equation}
    \label{eq:eta_turb}
    \eta_{\rm turb}\approx V_{A,y} (a/\lcs)^2_{\rm crit} \lcs.
\end{equation}
This corresponds to a turbulent Lundquist number $S_{\rm turb} = (\lcs/a)_{\rm crit}^2$ which, in the MHD case, would evaluate to $S_{\rm turb} \approx 10^4$, yielding the canonical plasmoid-mediated rate $\mathcal{R}=0.01$.
Naturally, a subgrid model would also prescribe a turbulent viscosity. 
From dimensional analysis alone, its value would be the same as that of $\eta_{\rm turb}$; direct numerical simulations are required to determine the value of the turbulent Prandtl number (although note that the numerical results of Ref.~\cite{loureiro_2012} hint that this may indeed be of order unity in the MHD case). 

For collisionless cases, there is not as much clarity on what the critical aspect ratio may be; figuring that out in different regions of parameter space is key for Eq.~(\ref{eq:eta_turb}) to be usable. 

\paragraph{Absence of scale separation.}
With the above considerations and conclusions in place, we can now inquire about the existence, or absence, of scale separation between the mean fields and the fluctuations.
Again borrowing insight from two-dimensional plasmoid chain dynamics, where the plasmoids' typical width in the transverse direction essentially defines the mean thickness of the layer, we similarly expect that in three-dimensional systems there will be no separation between the outer scale of the fluctuations and the scale of variation of the mean fields in the transverse direction, $a$; indeed, as already mentioned, we expect the width $a$ of variation of the mean fields to be defined by the fluctuations.
This is supported by existing numerical evidence~\cite{huang_2016,beresnyak_2017}.
We expect this to be true even in collisionless cases where kinetic-scale instabilities might be triggered, because those are expected to coexist with kink or tearing-type modes~\cite{alt_2019,winarto_2022} whose outer-scale in the $x$-direction will be $a$.

Consider now the out-of-plane-direction, with the background fields extending over a region of length $L_z$. 
Let us assume the (Alfv\'enic) turbulence in the sheet to be  critically balanced~\cite{goldreich_1995}, such that $\ell_\lambda/\lambda \sim V_{Az}/u_{x,\lambda}$, where $\lambda\le a$ is a lengthscale in the $x$-direction, and $\ell_\lambda$ and $u_{x,\lambda}$ are, respectively, the extent of the fluctuations in the direction parallel to the total field and the characteristic $x$-direction velocity fluctuation at that scale.
At the $x$-direction outer scale, then, we expect
\begin{align}
    \label{eq:z_ordering}
    \frac{\ell_a}{a}\sim \frac{V_{Az}}{u_{x,a}}\sim \frac{V_{Az}}{\Uin}\sim \frac{V_{Az}}{V_{Ay}}\frac{\lcs}{a}\sim \frac{L_z}{a},
\end{align}
where we have used Eq.~(\ref{eq:uin_mhd}) and assumed, in the last step, that the reconnecting fields are critically balanced, $\lcs/L_z\sim V_{Ay}/V_{Az}$. 
We thus find that, at the scale of the mean current sheet, there is no scale separation in the $z$ direction between the background and the fluctuations (note that for moderate guide fields 

Finally, one also expects $u_{x,\lambda}/\lambda \sim u_{y,\lambda}/\xi_\lambda$; evaluated at the outer scale, this relation implies $\xi_a/a\sim V_{Ay}/\Uin\sim \lcs/a$ and, thus, no scale separation between mean fields and fluctuations in the outflow direction.

These conclusions imply that  spatially averaging the fluctuations, instead of time averaging, may not be justified.

\paragraph{Computational procedure.}
The general description of the numerical approach to the solution of this problem is as follows.
Starting from prescribed functional forms for the background fields, one solves the fluctuating equations
until the necessary averaged quantities reach a steady-state. 
This solution is then fed into the mean-field equations, thus evolving said fields into a new state. 
This new mean state becomes the new background for the fluctuations, which will therefore adjust; and so on. 
The loop is repeated until the mean fields converge.

The different steps just described are non-trivial and require some consideration. 
Let us first consider the choice of the initial functional form of the mean fields. 
The most efficient prescription may be to specify a configuration that is supercritical, i.e., unstable to whatever instabilities are going to be driving the turbulence --- for example, in MHD one might initialize a  current sheet whose aspect ratio is set by the condition for triggering the plasmoid instability~\cite{loureiro_2007,loureiro_2013} in a forming sheet~\cite{Pucci_2014,uzdensky_2016,tenerani_2016,tolman_2018};
the (simplest) collisionless counterpart of this choice is a system-size Harris sheet~\cite{harris_1962} varying on the ion skin depth or smaller ~\cite{daughton_2011} (since, as argued above, such a sheet is unstable for realistic parameter choices). 
The ensuing dynamics that is captured by this numerical approach is then one of relaxation of the mean fields to a state where the effect of the turbulence that they drive is to balance the upstream inflow (see Fig.~\ref{fig:plots}).

One question which arises is that of whether a solution found through this procedure --- while obviously an admissible solution --- is, in fact, dynamically accessible. 
I.e., if the mean fields were to be dynamically evolved from a stable to an unstable configuration, would their final configuration be the same as that which results from this procedure? 
In the more complex cases, particularly when the layer is weakly collisional, there might be several instabilities possible, with different triggering conditions (e.g., some will depend on current gradients, others on the current itself, etc.) --- and, so, this procedure might converge to a solution that is governed by an instability that would never be triggered if the background mean fields had been evolved from an initially stable state. 
One way to address this problem is to perform a series of runs where the initial mean fields are made progressively less unstable (by decreasing the current gradient and/or the current itself) until a solution is found that is independent of the initial condition. 

Lastly, we remark on the different mesh sizes required for the solution of the mean fields and the fluctuations.
Whereas resolution of the full spectrum of fluctuations is required in the layer (but only there), the only requirement for the computation of the mean fields is to resolve the macroscopic, MHD-scale lengths $a$, $\lcs$ and $L_z$. 

\paragraph{Conclusions.}
In this Letter, we describe a framework for approaching reconnection as a transport-type problem, in which the evolution of the mean fields is self-consistently determined by the turbulence that they drive in the reconnection layer.
The validity of this method relies on the timescale separation between the fluctuations and the mean fields --- the former being Alfv\'enic or faster; the latter evolving on the reconnection timescale. 
It is argued that mean fields should remain on MHD scales even in the collisionless limit, a conclusion which has powerful implications for numerical modelling.
We describe a computational method to implement this approach whose efficiency gains over the brute force approach that has so far been used should be significant enough to enable three-dimensional reconnection simulations with asymptotically large scale separation between the system size and the micro-scales.\\

\begin{acknowledgments}
The author thanks Bill Daughton, Alex Schekochihin Adam Stanier and Dmitri Uzdensky for useful discussions. 
He is also thankful to the Center for Nonlinear Studies of the Los Alamos National Laboratory, where parts of this work were carried out.
Work supported by DOE award no. DE-SC0022012 and NSF-DOE Partnership in Basic Plasma Science and Engineering Award No. PHY-2010136.
\end{acknowledgments}

\bibliography{mean_field_rec}

\end{document}